\documentclass{jfm}
\usepackage{graphicx}
\usepackage{epstopdf, epsfig}

\usepackage{url}
\usepackage{amsmath}
\usepackage{amssymb}
\usepackage{graphicx}
\usepackage[svgnames]{xcolor}
\usepackage{physics}
\usepackage{amsmath}
\usepackage{mathrsfs}
\usepackage{soul}

\newcommand\qvec{\mathbf{q}}

\newcommand\xvec{\mathbf{x}}

\newcommand\Cmat{\mathbf{C}}

\newcommand\Mmat{\mathbf{M}}
\newcommand\Qmat{\mathbf{Q}}

\newcommand\Wmat{\mathbf{W}}

\newcommand\phivec{\boldsymbol{\phi}}

\newcommand\PhiMat{\boldsymbol{\Phi}}
\newcommand\LambdaMat{\boldsymbol{\Lambda}}

\newcommand{\qmean}{\overline{\qvec}}

\newcommand*{\vertbar}{\rule[-1ex]{0.5pt}{2.5ex}}

\shorttitle{Conditional space-time POD for intermittent and rare events}
\shortauthor{O. T. Schmidt and P. J. Schmid}

\title{{A conditional space-time POD formalism for
    intermittent and rare events: \\example of acoustic bursts in
    turbulent jets}}

\author{Oliver T. Schmidt\aff{1}
  \corresp{\email{oschmidt@ucsd.edu}}
 \and Peter J. Schmid\aff{2}}

\affiliation{\aff{1}Department of Mechanical and Aerospace
  Engineering, University of California San Diego, La Jolla, CA, USA
  \aff{2}Department of Mathematics, Imperial College London, London, UK}

\begin{document}

\maketitle

\begin{abstract}
  We present a conditional space-time proper orthogonal decomposition
  (POD) formulation that is tailored to the eduction of the average,
  rare or intermittent event from an ensemble of realizations of a
  fluid process. By construction, the resulting spatio-temporal modes
  are coherent in space and over a pre-defined finite time horizon and
  optimally capture the variance, or energy of the ensemble. For the
  example of intermittent acoustic radiation from a turbulent jet, we
  introduce a conditional expectation operator that focuses on the
  loudest events, as measured by a pressure probe in the far-field and
  contained in the tail of the pressure signal's probability
  distribution. Applied to high-fidelity simulation data, the method
  identifies a statistically significant `prototype', or average
  acoustic burst event that is tracked over time. Most notably, the
  burst event can be traced back to its precursor, which opens up the
  possibility of prediction of an imminent burst. We furthermore
  investigate the mechanism underlying the prototypical burst event
  using linear stability theory and find that its structure and
  evolution is accurately predicted by optimal transient growth
  theory. The jet-noise problem demonstrates that the conditional
  space-time POD formulation applies even for systems with probability
  distributions that are not heavy-tailed, i.e. for systems in which
  events overlap and occur in rapid succession.
\end{abstract}


\section{Introduction}

Intermittency is a ubiquitous feature that can be observed over a wide
range of dynamic systems arising in engineering and nature. Turbulent
shear flows, ocean waves, weather patterns, seismic events and
volcanic eruptions are but a few examples that are characterized by
the irregular occurrence of events far from the statistical
mean. Classically, intermittency is quantified in terms of a binary
intermittency function or an intermittency factor that describes the
probability of finding the flow in one state or the other. Once
identified, this factor can be used as a filter to obtain conditional
statistics for each state independently~\citep{Pope2000}. While these
statistics might accurately describe the random process in a
quantitative manner via its moments, the underlying physics are often
better understood from a dynamical systems perspective. In this light,
sporadic deviations from the statistical equilibrium can, for many
systems, be expressed as instabilities that are intermittently
triggered through external stochastic
excitation~\citep{mohamad2015probabilistic}. If the amplification
potential and growth rate of these instabilities are significant, a
statistical footprint in the process's distributional tails can be
observed and related to intermittent, rare events.

In the context of fluid mechanics, transient growth, associated with
the non-normality of the underlying linearized
dynamics~\citep{trefethen1993hydrodynamic,farrell1996generalized,schmid2001stability},
provides such an amplification mechanism. The linear theory that
governs transient growth allows for the computation of nonmodal
solutions that identify spatial structures that undergo significant
energy growth over short periods of time. Pipe flow is a prominent
example of a flow in which a transient linear amplification process
plays a crucial role as the flow transitions from the laminar to an
intermittently turbulent state with increasing flow
rate~\citep{kerswell2005recent,avila2011onset}. In particular, patches
of turbulence, so-called turbulent puffs, are observed despite the
fact that the laminar state is linearly stable to infinitesimal
disturbances. In other systems, nonlinear interactions may
sporadically conspire in such a way that responses of exceptionally
large amplitudes are observed. Such statistical outliers are commonly
referred to as extreme, or rare events due to their isolated
nature. An example of this class of intermittent effects are
oceanographic surface gravity waves of extreme magnitude, or rogue
waves~\citep{dysthe2008oceanic}.

\begin{figure}
  \centering
  \includegraphics[trim=0 0cm 0 0mm, clip,
    width=1\textwidth]{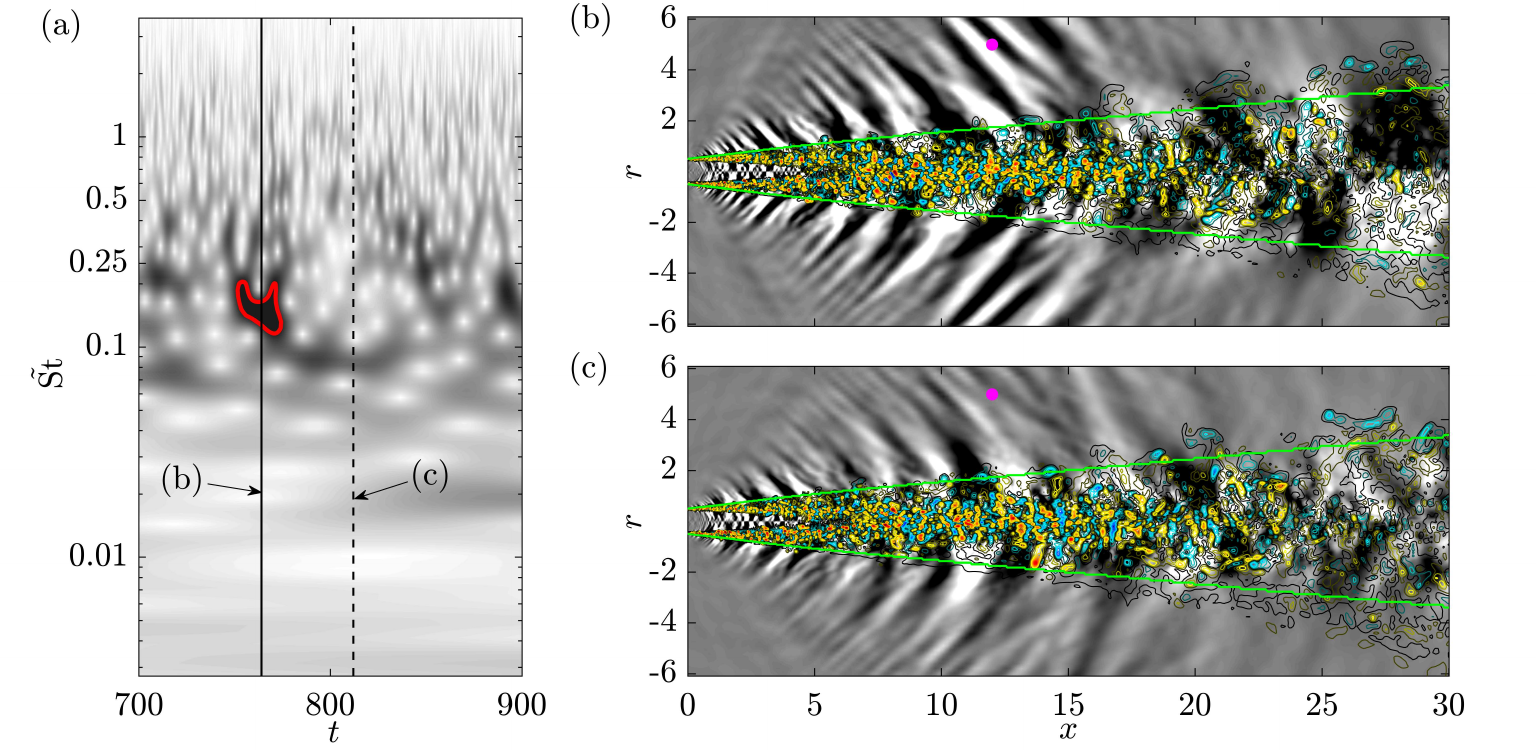}
  \caption{A high-energy burst in the acoustic far-field of a
    turbulent jet: (a) scaleogram of the $m=1$ component of the
    pressure signal at the probe location (magenta dot) at $(x_0, r_0)
    = (12, 5)$; (b,c) streamwise fluctuation velocity in the center
    plane (colored contours) and $m=1$ pressure component (grayscale
    contours) at $t$ = 366.8 in (b) and $t$ = 378 in (c). The two time
    instances represent the peak of the loud event identified in the
    scaleogram by the 75\% of maximum pressure contour (red) and the
    quiet period shortly thereafter. {The green lines in
      (b) and (c) indicate where the streamwise mean velocity
      corresponds to 10\% of the jet velocity,
      i.e. $U(\vb{x})=0.1U_j$, for later use in
      equation}~(\ref{eqn:weight_matrix}). A standard Morse wavelet is
    used in (a). All contour levels are adjusted for best
    readability.}
  \label{fig:LES_m0_burstVsQuiet}
\end{figure}

In this paper, we present a data-driven approach to isolate and
identify intermittent structures in a statistical manner by leveraging
both their temporal and spatial coherence. 
{The method builds on a general space-time POD formalism,
  as introduced in the original work of~\cite{Lumley:1970} and
  discussed in detail in \S~\ref{sec:problem}, to find structures that
  optimally represent the data in terms of variance, or energy, and
  combines it with the idea of conditional averaging to isolate a
  specific event from unconditional data. The use of conditional
  averages to study flow patterns dates back to pioneering work
  of~\cite{adrian1975role}, who introduced the method of linear
  stochastic estimation (LSE) to educe conditional flow structures
  from isotropic turbulence. LSE establishes a linear mapping between
  a conditional event and a conditionally-averaged flow variable. The
  mapping is computed from a minimization argument between the
  cross-correlation of the event data and the conditional flow state
  to be estimated. It is then used to compute conditionally-averaged
  estimates from unconditionally sampled two-point
  correlations. Stochastic estimation has been extended from single to
  multi-time estimates in the frequency
  domain~\citep{ewing1999examination,tinney2006spectral} and used in
  conjunction with POD~\citep{bonnet1994stochastic}. Unlike LSE, the
  present method does not rely on the \emph{a priori definition} of a
  candidate structure, the so-called `conditional eddy', but
  \emph{identifies} the energetically dominant structure by means of
  an eigenvalue decomposition of the conditional two-point space-time
  correlation tensor.} 
  
As an example, we choose the highly
intermittent acoustic Mach wave acoustic radiation of a hot turbulent
jet. The jet-noise problem is of eminent technical interest for the
aviation industry, in particular as it applies to naval aviation:
aircraft carrier flight crew personnel operate in one of the world's
loudest work environments, and regularly suffer from hearing loss due
to overexposure to intermittent
noise~\citep{aubert2011measurements}. In
figure~\ref{fig:LES_m0_burstVsQuiet}, we contrast two time instances
of the flow field obtained from a high-fidelity numerical
simulation~\citep{bres2017unstructured}. As
in~\cite{koenig2013farfield}, we use a wavelet transform of the
pressure signal at a probe location in the far-field to identify loud
events, or acoustic bursts, and distinguish them from quieter periods
in which the pressure fluctuations are closer to their statistical
mean.

A loud event is identified from the scaleogram in
figure~\ref{fig:LES_m0_burstVsQuiet}(a) and the corresponding
perturbation velocity and pressure fields for the $m=1$ azimuthal
Fourier component are shown in
figure~\ref{fig:LES_m0_burstVsQuiet}(b). The acoustic burst manifests
itself as a high-amplitude wave in the pressure field that is emitted
at a low angle (relative to the jet axis) from the jet. Shortly after,
the pressure field at the probe location appears quiet in
figure~\ref{fig:LES_m0_burstVsQuiet}(c). It is well established that
large-scale coherent structures, or wavepackets, are the dominant
source of low aft-angle jet noise and that these structures can be
modeled as spatial linear stability modes~\citep{jordan2013wave} or
resolvent modes~\citep{SchmidtEtAl_2018_JFM}. Their footprint can be
seen in the pressure field in the developing jet region with $x
\lesssim 10$ in
figure~\ref{fig:LES_m0_burstVsQuiet}(b,c). {A salient
  feature of wavepackets that is not represented by these modal
  solutions is their intermittent behavior, or jitter. Intimately
  linked to the jitter of the shear-layer instabilities, however, is
  the intermittency of the sound they radiate. The stochastic nature
  of jet noise has been measured}~\citep{juve1980intermittency,
  kearney2013intermittent}, quantified and
modeled~\citep{cavalieri2011jittering}, but is not fully understood as
of now.

In \S~\ref{sec:problem}, we first formulate the conditional space-time
POD problem that allows for the
inference of intermittent or rare events as space-time modes that
optimally capture the conditional variance of an ensemble of
realizations of the event. As an example, we then apply the general
framework to acoustic bursts in turbulent jets in \S~\ref{sec:jet},
before summarizing our findings in \S~\ref{sec:results}.

\section{A conditional space-time POD formulation}\label{sec:problem}

We start by defining the space-time inner product
\begin{equation}
  \langle \vb{q}_1,\vb{q}_2 \rangle_{\vb{x},t} =
  \int_{-\infty}^{\infty} \int_{V} \vb{q}_1^*(\vb{x},t)\, \Wmat(\vb{x}) \,
  \vb{q}_2(\vb{x},t) \, \dd V \, \dd t,
  \label{eqn:innerprod_spacetime}
\end{equation}
on which we also base a measure of perturbation size as the energy
(norm) of the quantity $\vb{q}.$ The diagonal, positive definite
weight matrix $\Wmat(\vb{x})$ is introduced to accommodate space
and/or variable dependent weights.
\begin{sloppypar}Equipped with the scalar product $\langle \cdot,\cdot \rangle_{\vb{x},t}$ and an
expectation operator $E\qty{\cdot}$, commonly defined as the sample
average, we embed ${\vb{q}(\vb{x},t)}$ into a Hilbert space
$\mathcal{H}$, which allows us to identify the spatio-temporal
structure $\phivec(\vb{x},t)\in\mathcal{H}$ that maximizes
\begin{equation}
  \lambda = \frac{E\qty{\qty|\left\langle
      \vb{q}(\vb{x},t),\phivec(\vb{x},t)
      \right\rangle_{\vb{x},t}|^2}}{\left\langle
    \phivec(\vb{x},t),\phivec(\vb{x},t)
    \right\rangle_{\vb{x},t}} \label{eqn:lambda_def0}
\end{equation}
using a variational approach. The $\phivec(\vb{x},t)$ and $\lambda$
that satisfy equation~(\ref{eqn:lambda_def0}) can be found as
solutions to the Fredholm eigenvalue problem
\begin{equation}\label{eqn:fredholm0}
  \int_{-\infty}^{\infty} \int_{V} \Cmat(\vb{x},\vb{x}',t,t')
  \phivec(\vb{x}',t') \, \dd \vb{x}' \, \dd t'\ = \lambda\,
  \phivec(\vb{x},t),
\end{equation}
where
$\Cmat(\vb{x},\vb{x}',t,t')=E\qty{\vb{q}(\vb{x},t)\vb{q}^*(\vb{x}',t')}$
denotes the two-point space-time correlation tensor. This formulation
corresponds to the classical space-time POD problem introduced
by~\cite{Lumley:1970}, which was vastly overshadowed by its popular
spatial variant~\cite[][]{sirovich1987turbulence, aubry1991hidden}. A
notable exception is the work by~\cite{gordeyev2013temporal}, in which
the authors solve a space-time POD problem to investigate flow
transients. More recently, the frequency-domain
version~\cite[see][]{towneschmidtcolonius_2018_jfm,
  SchmidtEtAl_2018_JFM}, which is derived from
equation~(\ref{eqn:fredholm0}) under the assumption of statistical
stationary, has gained popularity in the analysis of turbulent flows.

In deviation from other formulations, we introduce a conditional
expectation
\begin{equation}
  E\qty{ \qty| \left\langle \vb{q}(\vb{x},t),\phivec(\vb{x},t)
    \right\rangle_{\vb{x},t}|^2 \, \Big| \, H }
\end{equation}
with respect to the event $H.$ As we are interested in the average
evolution of the extreme event in space and time, we furthermore recast
the space-time inner product~(\ref{eqn:innerprod_spacetime}) into a
weighted space-time inner product
\begin{eqnarray}
  \langle \vb{q}_1,\vb{q}_2 \rangle_{\vb{x},\Delta T} &=& \int_{\Delta
    T} \int_{V} \vb{q}_1^*(\vb{x},t) \, \Wmat(\vb{x}) \,
  \vb{q}_2(\vb{x},t) \, \dd V \, \dd t,
  \label{eqn:innerprod_spacetime}
\end{eqnarray}
over some finite time horizon $\Delta T$ in the temporal neighborhood
of the extreme event $H$. The quantity to maximize
\begin{equation}
  \lambda = \frac{E\qty{\qty|\langle
      \vb{q}(\vb{x},t),\phivec(\vb{x},t) \rangle_{\vb{x},\Delta T}|^2
      \, \Big| \, H} }{\langle \phivec(\vb{x},t),\phivec(\vb{x},t)
    \rangle_{\vb{x},\Delta T}} \label{eqn:lambda_def},
\end{equation}
is defined analogously to equation~(\ref{eqn:lambda_def0}) and its
solution is obtained, analogous to equation~(\ref{eqn:fredholm0}),
from the corresponding weighted Fredholm eigenvalue problem
\begin{equation}\label{eqn:fredholm}
  \int_{\Delta T} \int_{V} \Cmat(\vb{x},\vb{x}',t,t') \Wmat(\vb{x}')
  \phivec(\vb{x}',t') \, \dd \vb{x}' \, \dd t'\ = \lambda
  \phivec(\vb{x},t).
\end{equation}
In discrete time and space, the eigenvalues and eigenvectors of the
two-point space-time correlation tensor are obtained from the
decomposition
\begin{equation}
  \label{eqn:evp0}
  \Qmat\Qmat^*\Mmat\PhiMat=\PhiMat\LambdaMat,
\end{equation}
where
\begin{eqnarray}\label{eqn:evp}
  \Qmat =
  \left[
    \begin{array}{cccc}
      \vertbar & \vertbar &        & \vertbar \\
      \vb{q}(t_0^{(1)}-t^-)    & \vb{q}(t_0^{(2)}-t^-)    & \ldots & \vb{q}(t_0^{(N_{\text{peaks}})}-t^-)    \\
      \vertbar & \vertbar &        & \vertbar \\
      \vdots   & \vdots  & \ddots & \vdots 	\\
      \vertbar & \vertbar &        & \vertbar \\
      \vb{q}(t_0^{(1)})    & \vb{q}(t_0^{(2)})    & \ldots & \vb{q}(t_0^{(N_{\text{peaks}})})    \\
      \vertbar & \vertbar &        & \vertbar	\\
      \vdots   & \vdots  & \ddots & \vdots 	\\
      \vertbar & \vertbar &        & \vertbar \\
      \vb{q}(t_0^{(1)}+t^+)    & \vb{q}(t_0^{(2)}+t^+)& \ldots     & \vb{q}(t_0^{(N_{\text{peaks}})}+t^+)    \\
      \vertbar & \vertbar &        & \vertbar
    \end{array}
    \right]
\end{eqnarray}
represents the space-time data matrix of realizations of events
occurring at times $t_0^{(j)}$. The $j$-th column of $\Qmat$ hence
contains the $j$-th realization of event $H$ evolving from time
$t_0^{(j)}-t^-$ to $t_0^{(j)}+t^+$, i.e.~over $\Delta T/\Delta t$ time
steps. Here, $\Delta t$ is the temporal separation between two
consecutive snapshots, and $t^-$ and $t^+$ define the time interval
\[
\left\{ t \left\vert t \in \qty[t_0^{(j)}-t^-, \, t_0^{(j)}+t^+]
\right. \right\}
\]
over which the $j$-th event evolves, with $\Delta T^{(j)}$ denoting
its length. The matrix $\Mmat$ accounts for both the weights
$\Wmat(\vb{x})$ and numerical quadrature weights stemming from the
spatial integration in equation~(\ref{eqn:fredholm}). The matrices
$\PhiMat = \left[\phi^{(1)}(\vb{x},t), \phi^{(2)}(\vb{x},t), \dots,
  \phi^{(N_{\text{peaks}})}(\vb{x},t) \right]$ and $\LambdaMat =
\hbox{diag}\left[\lambda^{(1)}, \lambda^{(2)}, \allowbreak \dots,
  \lambda^{(N_{\text{peaks}})} \right]$ contain the eigenvectors and
eigenvalues, i.e.~space-time POD modes and their corresponding
energies in the space-time inner product, respectively. In practice,
the large eigenvalue problem~(\ref{eqn:evp0}) is solved by computing
the cheaper eigendecomposition of the smaller matrix
$\Qmat^*\Mmat\Qmat$ first, in the same manner as for standard
POD~\citep{sirovich1987turbulence}. Note that the eigenvectors
$\phivec(\vb{x},t)$ of the space-time correlation tensor
{that constitute the conditional space-time POD modes are
  dependent on time. In contrast, classical and spectral POD modes are
  functions of space only.} By construction, space-time POD modes are
coherent in space and over the time horizon $\Delta T$, and orthogonal
in the space-time inner
product~(\ref{eqn:innerprod_spacetime}).\end{sloppypar}

\section{Example: Acoustic bursts in a round supersonic jet}\label{sec:jet}

As an example, we consider acoustic bursts that are intermittently
emitted from a high Reynolds number round jet and tailor the
space-time inner product~(\ref{eqn:innerprod_spacetime}) as well as
the event $H$ to this application. A total of $10,000$ snapshots
separated by $\Delta t = \Delta t^*c_\infty/D = 0.1$ acoustical time
units of a high-fidelity LES by~\cite{bres2017unstructured} of a
heated ideally expanded turbulent jet with jet Mach number $M_j =
U_j/c_j = 1.5$, Reynolds number $\Rey=\rho_jU_jD/\mu_j=155,000$ and
nozzle temperature ratio $T_0/T_\infty = 2.53$ are
analyzed. {A wall model was used to describe the turbulent
  boundary layer inside the nozzle (not part of the cylindrical domain
  considered below)}. The subscripts $j$, $\infty$ and $0$ indicate
jet, stagnation and freestream conditions, the superscript $*$ marks
dimensional quantities, $c$ is the speed of sound, $T$ the
temperature, $\mu$ the dynamic viscosity and $D$ the nozzle
diameter. The data for our study corresponds to the pressure field of
case \emph{A2} in~\cite{bres2017unstructured}, and the reader is
referred to that paper for more details. For cylindrical coordinates
with $\xvec=(x,r)$ and $\dd V=r \, \dd x \, \dd r$, where $x$ and $r$
are the streamwise and radial coordinates, respectively, the
rotational symmetry allows for a Fourier decomposition of the pressure
field
\begin{equation}\label{eqn:fft}
  \vb{p}(x,r,\theta,t) = \sum_{m}
  \hat{\vb{p}}_m(x,r,t)\mathrm{e}^{\mathrm{i}m\theta}
\end{equation}
into azimuthal Fourier modes $\hat{\vb{p}}_m(x,r,t)$. For our
demonstration, we focus on the $m=1$ component of the data as shown in
figure~\ref{fig:LES_m0_burstVsQuiet}, which contains the largest
fraction of the fluctuating energy of unheated
jets~\cite[][]{SchmidtEtAl_2018_JFM}. In the following, we drop the
$\hat{(\cdot)}_1$ with the understanding that $\vb{p}$ and $p$ denote
the $m=1$ Fourier component of the pressure. The pressure 2-norm
\[
\qty| \left \langle \vb{p}(\vb{x},t),\vb{p}(\vb{x},t) \right
\rangle_{\vb{x},\Delta T}|^2
\]
is chosen as a convenient measure for the acoustic energy of the flow.

In order to focus on the acoustic emissions into the far-field plus
the outer part of the shear-layer, we choose
\begin{equation}
  \Wmat(\vb{x}) = \begin{cases}
    1, & \text{if } \displaystyle{\frac{U(\vb{x})}{U_j}} \leq 0.1\\
    0, & \text{otherwise,}
  \end{cases} \label{eqn:weight_matrix}
\end{equation}
to mask out the highly energetic pressure fluctuations in the jet
column. Here, $U$ is the mean streamwise velocity and $U_j$ the jet
velocity. The threshold $U(\vb{x}) \leq 0.1 U_j$,
{i.e. the region outside the green lines in
  figure}~\ref{fig:LES_m0_burstVsQuiet}(b,c), is chosen as a suitable
way to distinguish these different parts of the flow. The inclusion of
the outer part of the shear-layer facilitates backtracking of the
acoustic burst event to its precursor in the mixing region close to
the nozzle. The final results were found qualitatively independent of
the exact choice of thresholding.

The event $H$ is specified to identify loud events, or acoustic
bursts, in the far-field of the jet. In particular, we detect such
events as local maxima in time from a single-point measurement of the
far-field pressure $p(\vb{x}_0,t)$ at some probe location
$\vb{x}_0 = (x_0,r_0)$. For the case at hand, we detect loud events at
$(x_0,r_0) = (12,5)$, see figure~\ref{fig:LES_m0_burstVsQuiet}. This
location corresponds to the peak location of the power spectral
density of the pressure along the edge of the domain, i.e., the
dominant aft-angle noise. A local maximum that identifies an event is
said to be detected at time $t_0$ if $|p(\vb{x}_0,t_0)|$ is larger than
its two neighboring samples at $t=t_0-\Delta t$ and $t=t_0+\Delta t$
in discrete time. The event is hence defined as
\begin{equation} \label{eqn:H}
  H : \qty{t_0^{(j)} \in t_{\text{sim}} \, \Big| \,
    |p(\vb{x}_0,t_0^{(j)}-\Delta t)| < |p(\vb{x}_0,t_0^{(j)})| >
    |p(\vb{x}_0,t_0^{(j)}+\Delta t)| },
\end{equation}
where $t_{\text{sim}} = [\Delta t,2\Delta t,\dots,10^4\Delta t]$ is
the set of discrete simulation times at which the LES snapshots have
been saved. We further restrict the conditional expectation
\begin{multline}
  E\qty(\left| \langle \cdot,\cdot \rangle_{\vb{x},\Delta T} \right|^2 \, \Big| \,
  H_{\text{peaks}}) \nonumber \\
  = \frac{1}{N_{\text{peaks}}}\sum_{j=1}^{N_{\text{peaks}}}\qty{
    \left| \langle \vb{p}(\vb{x},\Delta T^{(j)}),\phivec(\vb{x},\Delta
    T^{(j)}) \rangle_{\vb{x},\Delta T} \right|^2 \, \Big| \, H_{\text{peaks}}
  }, \\
  \text{with } H_{\text{peaks}} = \qty{H :
    |p(\vb{x}_0,t_0^{(1)})|\geq|p(\vb{x}_0,t_0^{(2)})|\geq\dots\geq|p(\vb{x}_0,t_0^{(N_{\text{peaks}})})|},
  \nonumber
\end{multline}
sorted such that the first $N_{\text{peaks}}$ largest peaks are
selected, i.e., we threshold the cardinality $|H|$ of the event.

\begin{figure}
  \centering
  \includegraphics[trim=0 0cm 0 0mm, clip,
    width=1\textwidth]{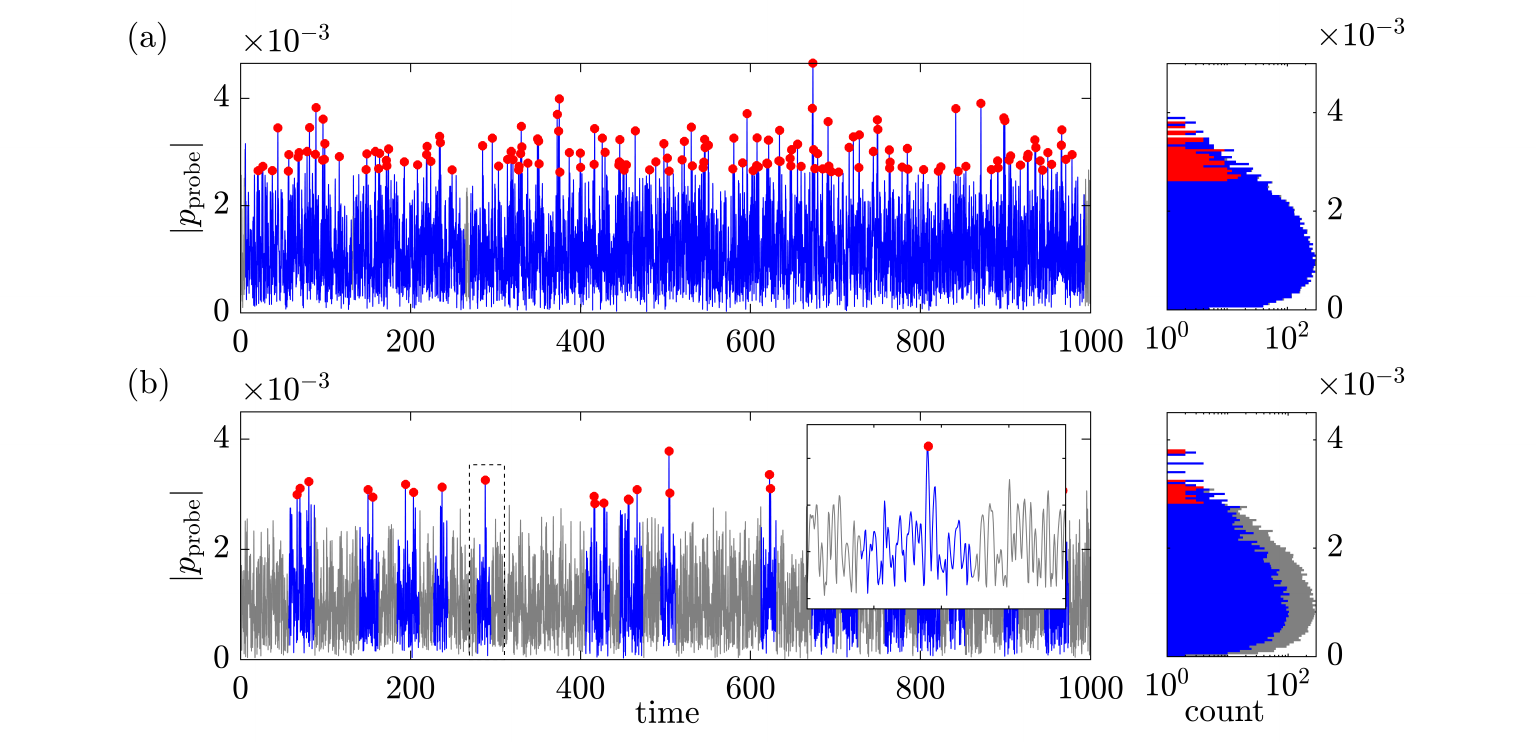}
  \caption{Time trace (left) and histogram (right) of the pressure at
    the probe location $(x_0,r_0) = (12,5)$: (a) $m=1$ component with
    $t^- = 150\Delta t$ and $t^+ = 149\Delta t$, such that $\Delta T =
    30$ spans $300$ snapshots centered about any $t_0^{(j)}$, and
    $N_{\text{peaks}} = 200$; (b) $m=1$ component with $t^- =
    100\Delta t$, $t^+ = 60\Delta t$, $N_{\text{peaks}} = 35$. The
    insert in (b) shows an example of an isolated event. The full
    signal is shown in gray; red symbols pinpoint events that occur at
    times $t_0^{(j)},$ with blue signals to indicate their temporal
    neighborhoods $t_0^{(j)}-t^- \leq \Delta T^{(j)} \leq
    t_0^{(j)}+t^+$.}
  \label{fig:ST-POD_pFFAndPDF_probeAndPeaks_2examples}
\end{figure}
Figure~\ref{fig:ST-POD_pFFAndPDF_probeAndPeaks_2examples} shows the
time trace of the pressure signal at the probe location and its
histogram for two different choices of space-time POD parameters
$t^-$, $t^+$, and $N_{\text{peaks}}$. The parameter combination $t^- =
150$, $t^+ = 149$, and $N_{\text{peaks}} = 200$ shown in
figure~\ref{fig:ST-POD_pFFAndPDF_probeAndPeaks_2examples}a is used in
the remainder of this paper. This choice centers the extreme event within
a time interval of $30$ acoustic units, which approximates the time it
takes an acoustic pulse to transverse the domain. The combination
shown in figure~\ref{fig:ST-POD_pFFAndPDF_probeAndPeaks_2examples}b
solely serves as a less dense example, with the insert highlighting a
single isolated event (filled red circle) occurring at $t_0 = 288.1$
and its temporal neighborhood, or time segment, $278.1 \leq \Delta T
\leq 294.6$ (blue line segment). Note that every snapshot might well
contain multiple extreme events at different stages of their temporal
evolution which results in overlapping time segments. The plot shown
on the right side of
figure~\ref{fig:ST-POD_pFFAndPDF_probeAndPeaks_2examples} uses the
same color representation as used for the time trace to distinguish
the histogram of the entire signal (gray) from those of the event
(red) and segments (blue). As we are considering the absolute value of
a presumingly Gaussian-distributed random signal, the time trace
histogram can be approximated by a $\chi^2$-distribution. By the
definition of $H$ in equation~(\ref{eqn:H}), the extreme acoustic events
occupy the tail of this distribution.

\begin{figure}
  \centering
  \includegraphics[trim=0 0cm 0 0mm, clip,
    width=1\textwidth]{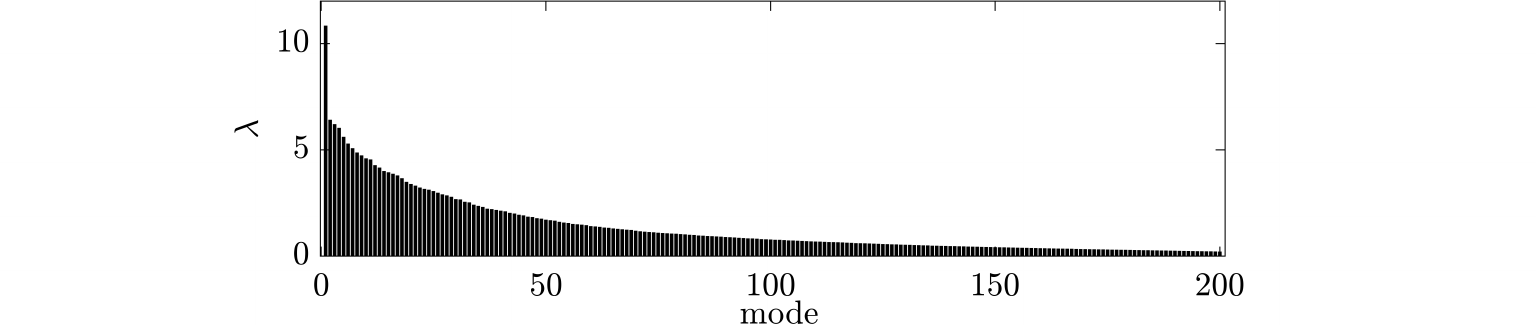}
  \caption{Conditional space-time POD energy spectrum for $t^- =
    150\Delta t$, $t^+ = 149\Delta t$, and $N_{\text{peaks}}=200$.}
  \label{fig:ST-POD_energySpectrum}
\end{figure}
In figure~\ref{fig:ST-POD_energySpectrum}, we present the eigenvalue,
or energy spectrum of the conditional space-time POD problem defined
by the parameters presented in
figure~\ref{fig:ST-POD_pFFAndPDF_probeAndPeaks_2examples}(a). Two
features of the spectrum stand out: (i) the dominance of the principal
eigenvalue which is $\approx 40\%$ larger than the following
eigenvalue; (ii) the slow decay of the rest of the spectrum starting
with the second mode. These two observations are crucial to our
analysis as they indicate that the particular space-time structure
associated with the first eigenvalue plays a dominant role in the
space-time statistics.


\begin{figure}
  \centering
  \includegraphics[trim=0 32mm 0 5mm, clip,
    width=1\textwidth]{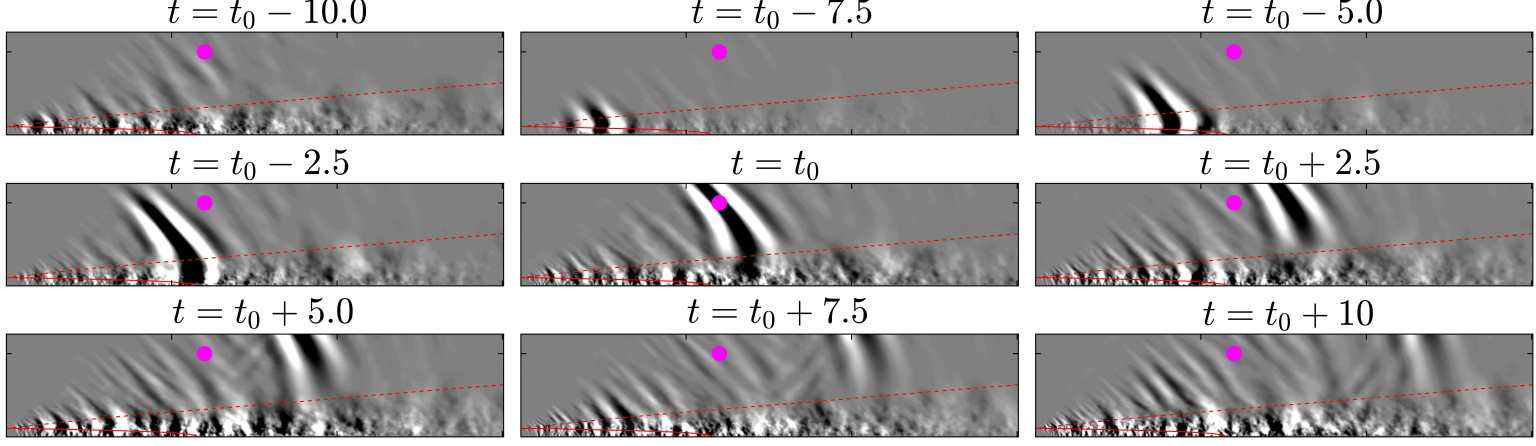}

  \caption{Temporal evolution of the leading conditional space-time
    POD mode
    ($\blacksquare\!{\textcolor{gray}{\blacksquare}}\!\square$,
    $-0.25<\phi^{(1)}(\vb{x},t)/\|\phi^{(1)}(\vb{x},t)\|_\infty<0.25$). The
    potential core and the jet width are indicated as lines of
    constant streamwise mean velocity $U$ at 99\% (solid red) and 5\%
    (dashed red) of the jet velocity $U_j$, respectively. The
    space-time integral energy of this mode is given by the leading
    eigenvalue $\lambda^{(1)}$ in
    figure~\ref{fig:ST-POD_energySpectrum}. An animation of this mode
    in direct comparison with the linear optimal solution (shown in
    figure~\ref{fig:optMode_tilePlot} below) can be found in the
    supplementary material.}
  \label{fig:ST-POD_tilePlot}
\end{figure}
\begin{sloppypar} We next investigate the corresponding structure, i.e., the leading
space-time POD mode $\phi^{(1)}(\vb{x},t)$, in
figure~\ref{fig:ST-POD_tilePlot}. Nine representative time instances
$t^{(j)} = \qty[50\Delta T, 75\Delta T, 100\Delta T, \dots, 250\Delta
  T]$ that track the event's evolution from its formation in the
near-nozzle shear-layer (top left) to the time it exits the
computational domain (lower right) are shown. The first instance at $t
= t_0-10$ represents the earliest time for which we observe a distinct
localized wavepacket structure. This structure is identified as the
initial seed, or precursor, of the acoustic burst, and over time
evolves into an acoustic burst with a distinct wavenumber, spatial
extent and ejection angle. The existence of a statistically prevalent
mode was \emph{a priori} not obvious, and suggests the existence of an
underlying physical mechanism. \end{sloppypar}

Motivated by a number of studies that demonstrate that mean flow
stability analysis can yield accurate predictions of the dynamics of
turbulent flows (see, e.g.,~\cite{beneddine2016conditions}), we seek a
description of the average linear burst event in terms of its linear
dynamics, here represented by the compressible Navier-Stokes operator
$\mathsfbi{A} = \mathsfbi{A}(\qmean,\vb{x},m)$ linearized about the
mean state $\qmean$. {In absence of a model for the
  effective Reynolds number in mean flow stability analyses and as
  in}~\cite{SchmidtEtAl_2018_JFM}, { we let
  $\Rey=3\times10^4$. The reader is referred to the latter reference
  for a more detailed discussion and for details of the
  discretization, which omits the nozzle.}
The optimality of the conditional POD formulation in the space-time
inner product leads itself to an optimal transient growth problem

\begin{equation}
  G(\tau) = \max_{\vb{q}(0)} \frac{\|
      \vb{q}(\tau)\|_E^2}{\|
      \vb{q}(0)\|_E^2}
    = \max_{\vb{q}(0)}\frac{\langle
      \mathscr{A}(\tau)\vb{q}(0),\mathscr{A}(\tau)\vb{q}(0)
      \rangle_E}{\langle
    \vb{q}(0),\vb{q}(0)
    \rangle_E}
        = \max_{\vb{q}(0)}\frac{\langle
      \vb{q}(0),\mathscr{A}^\dagger(\tau)\mathscr{A}(\tau)\vb{q}(0)
      \rangle_E}{\langle
    \vb{q}(0),\vb{q}(0)
    \rangle_E}\label{eqn:opt_growth}
\end{equation}
that maximizes the growth of energy as measured in the spatial inner
product $\langle \vb{q}_1,\vb{q}_2 \rangle_E = \int_{V}
\vb{q}_1^*(\vb{x})\, \Wmat'(\vb{x})\, \vb{q}_2(\vb{x}) \, \dd V$ and
over a finite time $\tau = t^- + t^+$. Here, $\Wmat'(\vb{x})$ contains
the same spatial restriction given by
equation~(\ref{eqn:weight_matrix}) as before {and, in
  addition, the component-wise weights of the compressible energy norm
  first introduced by}~\cite{chu1965energy}.

$\mathscr{A}(\tau) = e^{\mathsfbi{A}\tau}$ represents the evolution
operator that maps $\vb{q}(0)$ onto $\vb{q}(\tau)$, and
$\mathscr{A}^\dagger(\tau)=e^{\mathsfbi{A}^*\tau}$ is defined as its
adjoint counterpart {that maps $\vb{q}^\dagger(\tau)$ onto
  $\vb{q}^\dagger(0)$ back in time}. As in~\cite{schmidt2015optimal},
we converge the initial condition $\vb{q}(0)$ by direct-adjoint
looping, and enforce the spatial restriction implied by
equation~(\ref{eqn:weight_matrix}) at the beginning of each
iteration. A comprehensive review of nonmodal theory can be found
in~\cite{2007AnRFM..39..129S}.

\begin{figure}
  \centering
  \includegraphics[trim=0 32mm 0 3mm, clip,
    width=1\textwidth]{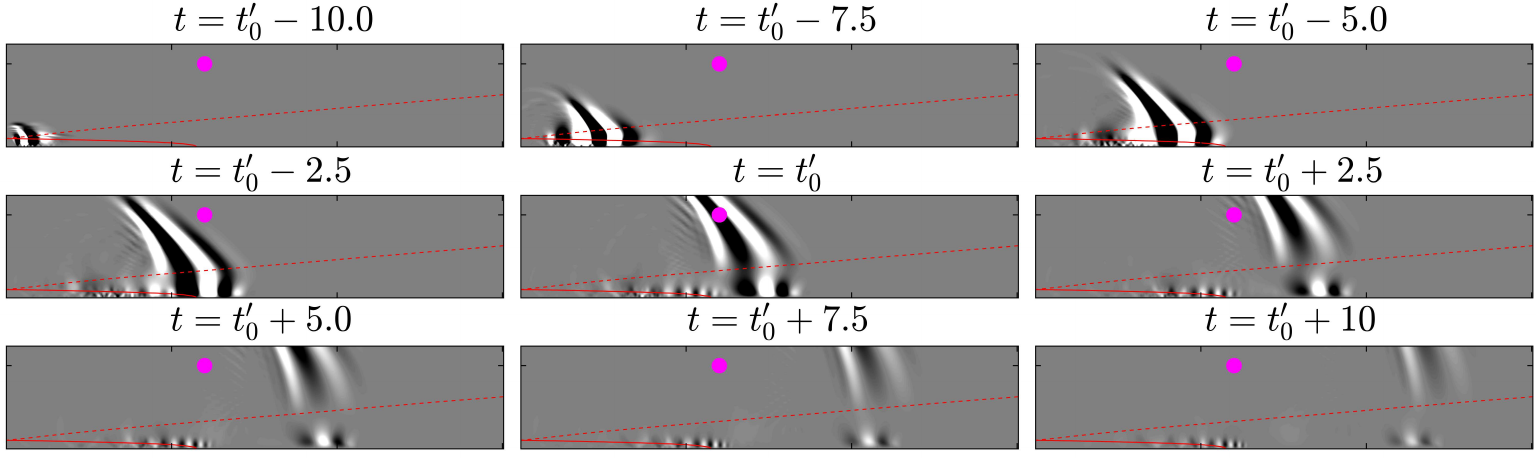}
  \caption{Same as figure~\ref{fig:ST-POD_tilePlot}, but for the
    pressure component of the optimal initial condition $\vb{q}(0)$
    ($\blacksquare\!{\textcolor{gray}{\blacksquare}}\!\square$,
    $-0.25<p(t)/\|p(t)\|_\infty<0.25$). This optimal solution should
    be compared to the leading conditional space-time POD mode
    depicted in figure~\ref{fig:ST-POD_tilePlot}.}
  \label{fig:optMode_tilePlot}
\end{figure}

The time evolution of the pressure component of the optimal initial
condition $\vb{q}(0)$ is reported in
figure~\ref{fig:optMode_tilePlot}. The gain associated with this
optimal structure, as defined in equation~(\ref{eqn:opt_growth}), is
$G(\tau)=231$. The waves in the end region of the potential core are
described in detail by~\cite{tam1989three}
and~\cite{TowneEtAl_2017_JFM}, and are not relevant for this
study. The time offset $t'_0 = 40$ aligns the evolution of the optimal
structure with that of the leading conditional space-time POD mode
depicted in figure~\ref{fig:ST-POD_tilePlot}, and is introduced
without loss of generality as we are interested in the evolution
relative to the occurrence of an event at the probe location. A
remarkable correspondence between the optimal structure and the
dominant space-time POD mode is observed. This correspondence strongly
suggests that nonmodal, transient growth plays a critical role in the
generation of the average acoustic burst event. {The close
  correspondence at early times in particular, suggests that the
  precursor takes the form of the optimal initial condition, which in
  turn corresponds to the adjoint solution
  $\vb{q}^\dagger(0)=\vb{q}(0)$. This suggests that an acoustic burst
  is triggered whenever turbulent fluctuations randomly align with the
  theoretical optimal initial condition, i.e.~if their projection onto
  the adjoint is large. In this scenario, the associated gain is then
  more appropriately interpreted as a sensitivity of
  output-perturbations with appreciable amplitudes to small-amplitude
  initial disturbances.}

\section{Conclusions}\label{sec:results}

We present a data-driven approach based on a conditional space-time
POD formulation that is tailored to the eduction and statistical
description of intermittent or rare events from data. The resulting
modes are orthonormal in a space-time inner product and optimally
capture the conditional variance of an ensemble of realizations of the
intermittent event. By construction, the modes are coherent in space
and over a finite time horizon. The formalism hence bridges the gap
between standard POD modes, which are coherent at zero time lag only,
and spectral POD modes, which are perfectly coherent over all time.

It can also be thought of as an extension to classical time-resolved
conditional averaging that has previously been applied as a structure
identification technique, for example to identify acoustic source
locations in jets~\citep{guj2003acoustic}, or more recently to examine
the hairpin generation process in turbulent wall boundary
layers~\citep{hack2018coherent}. Compared to conditional averaging,
the method introduced here has the additional advantage that it
inherits the desirable mathematical properties from the POD formalism.

To demonstrate the approach, we apply conditional space-time POD to
the example of Mach wave radiation of a turbulent jet. We find a
statistically dominant space-time mode that is energetically well
separated from all other modes. It takes the form of a spatially
confined wavepacket that originates in the thin initial shear-layer
close to the nozzle. Over time, this initial seed evolves into an
acoustic burst of a distinct wavenumber and spatial support that is
emitted into the far-field. The structure and dynamics of this
prototype acoustic burst suggest the existence of an underlying
physical mechanism. We investigate the possibility of optimal
transient growth as a candidate mechanism, and find a remarkable
agreement between the evolution of the optimal initial condition and
the leading POD mode.

The orthogonality and optimality properties of the conditional
space-time POD modes make them potential candidates for basis
functions for use in reduced-order models of transient processes. In
fluid mechanics, Galerkin models~\citep[see, e.g.,][]{book:760714} are
widely successful and build on exactly these properties. The temporal
coherence of the modes furthermore enables the identification of
precursors of extreme events, such as the initial seed of the acoustic
burst event shown in the upper-left panel of
figure~\ref{fig:ST-POD_tilePlot}. We plan to leverage this capability
for model-predictive control in the future.

\paragraph{\bf{Acknowledgements}}

We gratefully acknowledge support during the 2018 Summer Program at
the Center for Turbulence Research at Stanford University where this
work was initiated. O.T.S.~thanks Tim Colonius and gratefully
acknowledges support from the Office of Naval Research under grant
No. N00014-16-1-2445 with Dr. Knox Millsaps as program manager for the
portion of the work that was completed at Caltech. Thanks are due to
Aaron Towne, Sanjiva Lele and Stefan Llewellyn Smith for fruitful discussions and valuable
comments. The LES study was supported by the NAVAIR SBIR project,
under the supervision of Dr. John T. Spyropoulos. The main LES
calculations were carried out on CRAY XE6 machines at DoD HPC
facilities in ERDC DSRC.

\bibliographystyle{jfm}

\bibliography{jets,acoustics,others,non-modalStab,mypublications_proceedings,mypublications_journals,books,PSE_and_LST,PODetc,stochastics}

\end{document}